\documentstyle[12pt,epsfig]{article}

\def\myfigure#1{#1}

\def\cA{{\cal A}}
\def\cP{{\cal P}}

\newcommand{\rms}{\rm\scriptsize}
\def\NPPS{Nucl. Phys. B (Proc. Suppl.)}
\def\PL{Phys. Lett.}
\def\PR{Phys. Rev.}
\def\NP{Nucl. Phys.}
\def\ZP{Z. Phys.}

\def\beq{\begin{equation}}
\def\eeq#1{\label{#1}\end{equation}}
\def\eeqn{\end{equation}}

\def\beqa{\begin{eqnarray}}
\def\eeqa#1{\label{#1}\end{eqnarray}}
\def\eeqan{\end{eqnarray}}
\def\CR{\nonumber \\ }

\def\leqn#1{(\ref{#1})}

\def\Journal#1#2#3#4{{#1} {\bf #2}, #3 (#4)}
\def\add#1#2#3{{\bf #1}, #2 (#3)}

\def\PRL{Phys. Rev. Lett.}

\def\PR{Phys. Rev.}

\let\bar=\overbar

\def\etal{{\it et al.}}

\def\O{{\cal O}}

\def\Dslash{\not{\hbox{\kern-4pt $D$}}}
\def\dslash{\not{\hbox{\kern-2pt $\del$}}}

\def\msb{{\bar{\ssstyle M \kern -1pt S}}}

\def\Title#1{\begin{center} {\Large {\bf #1} } \end{center}}

\begin{document}

\Title{Tau Physics}

\bigskip\bigskip


\begin{raggedright}  

{\it A. Pich\index{Pich, A.}\\  IFIC,
Universitat de Val\`encia --- CSIC\\
Apt. Correus 2085, E-46071 Val\`encia, Spain}
\bigskip\bigskip
\end{raggedright}

\section{Introduction}
\label{sec:introduction}

The $\tau$ lepton is a member of
the third generation which decays into particles belonging to the first
and second ones.
Thus, $\tau$ physics could provide some
clues to the puzzle of the recurring families of leptons and quarks.
One na\"{\i}vely expects the heavier fermions to be more sensitive to
whatever dynamics is responsible for the fermion--mass generation.
The pure leptonic or semileptonic character of $\tau$  decays
provides a clean laboratory to test the structure of the weak
currents  and the universality of their couplings to the gauge bosons.
Moreover, the  $\tau$ is
the only known lepton massive enough to  decay  into  hadrons;
its  semileptonic decays are then an ideal tool for studying
strong interaction effects in  very clean conditions.

 The last few years have witnessed a substantial change on our knowledge
of the $\tau$ properties \cite{tau98,taurev}.
The large (and clean) data samples collected by the most recent experiments
have improved considerably the statistical accuracy and, moreover,
have brought a new level of systematic understanding.

\section{Charged--Current Universality}
\label{sec:cc}

The decays   
$\tau^-\to e^-\bar\nu_e\nu_\tau$ and $\tau^-\to\mu^-\bar\nu_\mu\nu_\tau$
are theoretically understood at the level of the electroweak
radiative corrections \cite{MS:88}. Within the Standard Model (SM),
\begin{equation}
\label{eq:leptonic}
\Gamma (\tau^- \rightarrow \nu_{\tau} l^- \bar{\nu}_l)  \, = \,
  {G_F^2 m_{\tau}^5 \over 192 \pi^3} \, f(m_l^2 / m_{\tau}^2) \, 
r_{EW},
\end{equation}
where 
$f(x) = 1 - 8 x + 8 x^3 - x^4 - 12 x^2 \log{x}$.
The factor $r_{EW}=0.9960$ takes into account radiative corrections 
not included in the
\index{Fermi coupling constant}
Fermi coupling constant $G_F$, and the non-local structure of the
$W$ propagator \cite{MS:88}.
Using the value of $G_F$ measured in $\mu$ decay, 
$G_F = (1.16637 \pm 0.00001)\times 10^{-5}\:\mbox{\rm GeV}^{-2}$
 \cite{MS:88,RS:99},
Eq.~\leqn{eq:leptonic} 
provides a relation between the $\tau$ lifetime
and the leptonic branching ratios
$B_{\tau\to l}\equiv B(\tau^-\to\nu_\tau l^-\bar\nu_l)$:
\beq
\label{eq:relation}
B_{\tau\to e} \, = \, {B_{\tau\to\mu} \over 0.972564\pm 0.000010} 
\, = \,
{ \tau_{\tau} \over (1.6321 \pm 0.0014) \times 10^{-12}\, {\rm s} } \, .
\eeqn
The errors reflect the present uncertainty of $0.3$ MeV
in the value of $m_\tau$ \cite{BES,PDG:98}.

\goodbreak

\begin{table}[thb]
\begin{center}
\begin{tabular}{l|r}
\hline 
$m_\tau$ & $(1777.05^{+0.29}_{-0.26})$ MeV \\
$\tau_\tau$ & $(290.77\pm 0.99)$ fs \\
Br($\tau^-\to\nu_\tau e^-\bar\nu_e$) & $(17.791\pm 0.054)\% $ \\
Br($\tau^-\to\nu_\tau\mu^-\bar\nu_\mu$) & $(17.333\pm 0.054)\% $ \\
Br($\tau^-\to\nu_\tau\pi^-$) & $(11.02\pm 0.09)\% $ \\
Br($\tau^-\to\nu_\tau K^-$) & $(0.690\pm 0.025)\% $ \\
\hline
\end{tabular}
\caption{World average values 
for some basic $\tau$ parameters.}
\label{tab:parameters}
\end{center}
\end{table}
%

\begin{figure}[htb]
\begin{center}
\myfigure{\epsfxsize =10cm \epsfbox{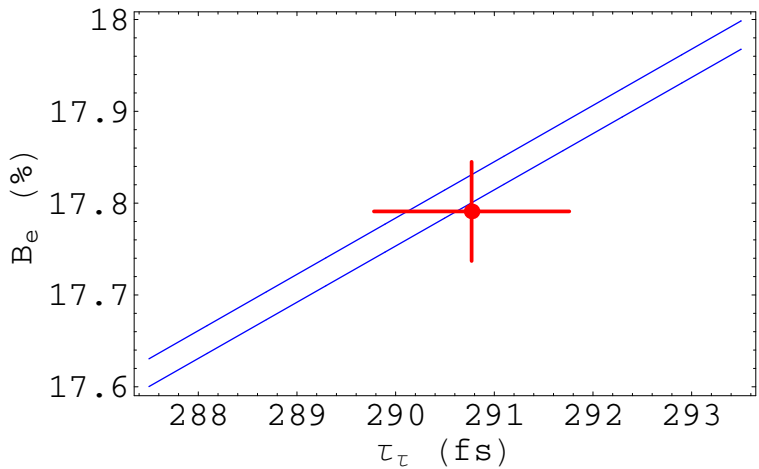}}
\caption{Relation between $B_{\tau\to e}$ and $\tau_\tau$. The
band corresponds to Eq.~(\protect\ref{eq:relation}).
\label{fig:BeLife}}
\end{center}
\end{figure}

The relevant experimental measurements are given in Table~\ref{tab:parameters}.
The predicted $B_{\tau\to\mu}/B_{\tau\to e}$ ratio is in perfect
agreement with the measured value 
$B_{\tau\to\mu}/B_{\tau\to e} = 0.974 \pm 0.004$.  
As shown in Fig.~\ref{fig:BeLife}, the relation between
$B_{\tau\to e}$ and
$\tau_\tau$ is also well satisfied by the present data. 
The experimental precision (0.3\%) is already approaching the
level where a possible non-zero $\nu_\tau$ mass could become relevant; the
present bound \cite{ALEPHnumass}
$m_{\nu_\tau}< 18.2$ MeV (95\% CL) only guarantees that such effect
is below 0.08\%.

These measurements can be used to test the universality of
the $W$ couplings to the leptonic charged currents.
The ratio $B_{\tau\to\mu}/B_{\tau\to e}$ constrains $|g_\mu/g_e|$,
while
$B_{\tau\to e}/\tau_\tau$ and  $B_{\tau\to\mu}/\tau_\tau$
provide information on $|g_\tau/g_\mu|$ and $|g_\tau/g_e|$.
The present results are shown in Table~\ref{tab:universality},
together with the values obtained from the ratios    
$\Gamma(\pi^-\to e^-\bar\nu_e)/\Gamma(\pi^-\to\mu^-\bar\nu_\mu)$ 
\cite{BR:92} and
$\Gamma(\tau^-\to\nu_\tau P^-)/
\Gamma(P^-\to \mu^-\bar\nu_\mu)$ \  [$P=\pi,K$].
Also shown are the constraints obtained from the
$W^-\to l^-\bar\nu_l$ decay modes at the $p$-$\bar p$ colliders
\cite{PDG:98,EL:99} and LEP II \cite{LEP:99}.
The present data verify the universality of the leptonic
charged--current couplings to the 0.15\% ($e/\mu$) and 0.23\%
($\tau/\mu$, $\tau/e$) level. 

\begin{table}[htb]
\begin{center}
\begin{tabular}{l|c|c|c}
& $|g_\mu/g_e|$  & $|g_\tau/g_\mu|$ & $|g_\tau/g_e|$ \\ \hline
$B_{\tau\to\mu}/B_{\tau\to e}$   & $1.0009\pm 0.0022$ & --- & --- 
\\
$B_{\tau\to e}\,\,\tau_\mu/\tau_\tau$ & ---  & $0.9993\pm 0.0023$ & --- 
\\
$B_{\tau\to\mu}\,\,\tau_\mu/\tau_\tau$ & ---  & ---  & $1.0002\pm 0.0023$
\\
$B_{\pi\to e}/B_{\pi\to\mu}$  & $1.0017\pm 0.0015$  &  --- & --- 
\\
$\Gamma_{\tau\to\pi}/\Gamma_{\pi\to\mu}$ & ---  & $1.005\pm 0.005$ & --- 
\\
$\Gamma_{\tau\to K}/\Gamma_{K\to\mu}$ & ---  & $0.981\pm 0.018$ & --- 
\\
$B_{W\to l}/B_{W\to l'}$ \  ($p\bar p$) & $0.98\pm 0.03$ &  --- &
$0.987\pm 0.025$
\\
$B_{W\to l}/B_{W\to l'}$ \ (LEP2) & $1.002\pm 0.016$ & 
$1.008\pm 0.019$ & $1.010\pm 0.019$
\\ \hline
\end{tabular}
\caption{Present constraints on charged--current lepton
universality.}
\label{tab:universality}
\end{center}
\end{table}

\section{Neutral--Current Universality}
\label{sec:nc}

In the SM, all leptons with equal electric charge have identical
couplings to the $Z$ boson.
This has been tested at LEP and SLC \cite{LEP:99},
by measuring the total $e^+e^-\to Z \to l^+l^-$
cross--section, the forward--backward asymmetry,
the (final) polarization asymmetry, the forward--backward (final) polarization
asymmetry, and (at SLC) the left--right asymmetry between the
cross--sections for initial left-- and right--handed electrons
and the left--right forward--backward asymmetry.
$\Gamma_l\equiv\Gamma(Z\to l^+l^-)$ 
determines the sum $(v_l^2 + a_l^2)$,
where $v_l$ and $a_l$ are the effective vector and axial--vector $Z$ 
couplings, while the ratio $v_l/a_l$ 
is derived from the asymmetries which measure the average longitudinal 
polarization of the lepton $l^-$:
\beq
\cP_l \, \equiv \, { - 2 v_l a_l \over v_l^2 + a_l^2} \, .
\eeq{eq:P_l}
The measurement of the final polarization asymmetries can (only) be done for 
$l=\tau$, because the spin polarization of the $\tau$'s
is reflected in the distorted distribution of their decay products.
Thus, $\cP_\tau$ and $\cP_e$ can be determined from a
measurement of the spectrum of the final charged particles in the
decay of one $\tau$, or by studying the correlated distributions
between the final decay products of both $\tau's$ \cite{ABGPR:92}.

\begin{table}[thb]
\begin{center}
\begin{tabular}{c|c|c|c|c}
& $e$ & $\mu$ & $\tau$ & $l$ 
\\ \hline
$\Gamma_l$ \, (MeV) & $83.90\pm 0.12$ & $83.96\pm 0.18$ & $84.05\pm 0.22$ & 
$83.96\pm 0.09$
\\
$\cA_{\mbox{\rms FB}}^{0,l}$ \, (\%) & $1.45\pm 0.24$ & $1.67\pm 0.13$ & 
$1.88\pm 0.17$ & $1.701\pm 0.095$
\\ \hline
\end{tabular}
\caption{Measured values \protect\cite{LEP:99,swartz} of 
$\Gamma_l$    and   $\cA_{\mbox{\protect\rms FB}}^{0,l}$.
The last column shows the combined result 
(for a massless lepton) assuming lepton universality.}
\label{tab:LEP_asym}
%
\vspace{0.75cm}
\begin{tabular}{c|c||c|c}
\hline
$-\cA_{\mbox{\rms Pol}}^{0,\tau} = -\cP_\tau$ &
$0.1425\pm 0.0044$ & 
${4\over 3}\cA^{0,e}_{\mbox{\rms FB,LR}} \Rightarrow -\cP_e$ &
$0.1558\pm 0.0064$
\\
$-{4\over 3}\cA^{0,\tau}_{\mbox{\rms FB,Pol}} = -\cP_e$ &
$0.1483\pm 0.0051$ & 
${4\over 3}\cA^{0,\mu}_{\mbox{\rms FB,LR}} = -\cP_\mu$ &
$0.137\pm 0.016$
\\
$\cA_{\mbox{\rms LR}}^0 = -\cP_e$ &
$0.1511\pm 0.0022$ &
${4\over 3}\cA^{0,\tau}_{\mbox{\rms FB,LR}} = -\cP_\tau$ &
$0.142\pm 0.016$
\\
$\{{4\over 3}\cA_{\mbox{\rms FB}}^{0,l}\}^{1/2} = -P_l$ &
$0.1506\pm 0.0042$ & & 
\\ \hline
\end{tabular}
\caption{$\cP_l$ determinations from different asymmetry measurements
\protect\cite{LEP:99,swartz,SLD:99}.}
\label{tab:pol_asym}
\end{center}
\end{table}
%

\begin{figure}[thb]
\centerline{
\begin{minipage}[c]{.5\linewidth}
\begin{center}
\myfigure{\epsfxsize =7cm \epsfbox{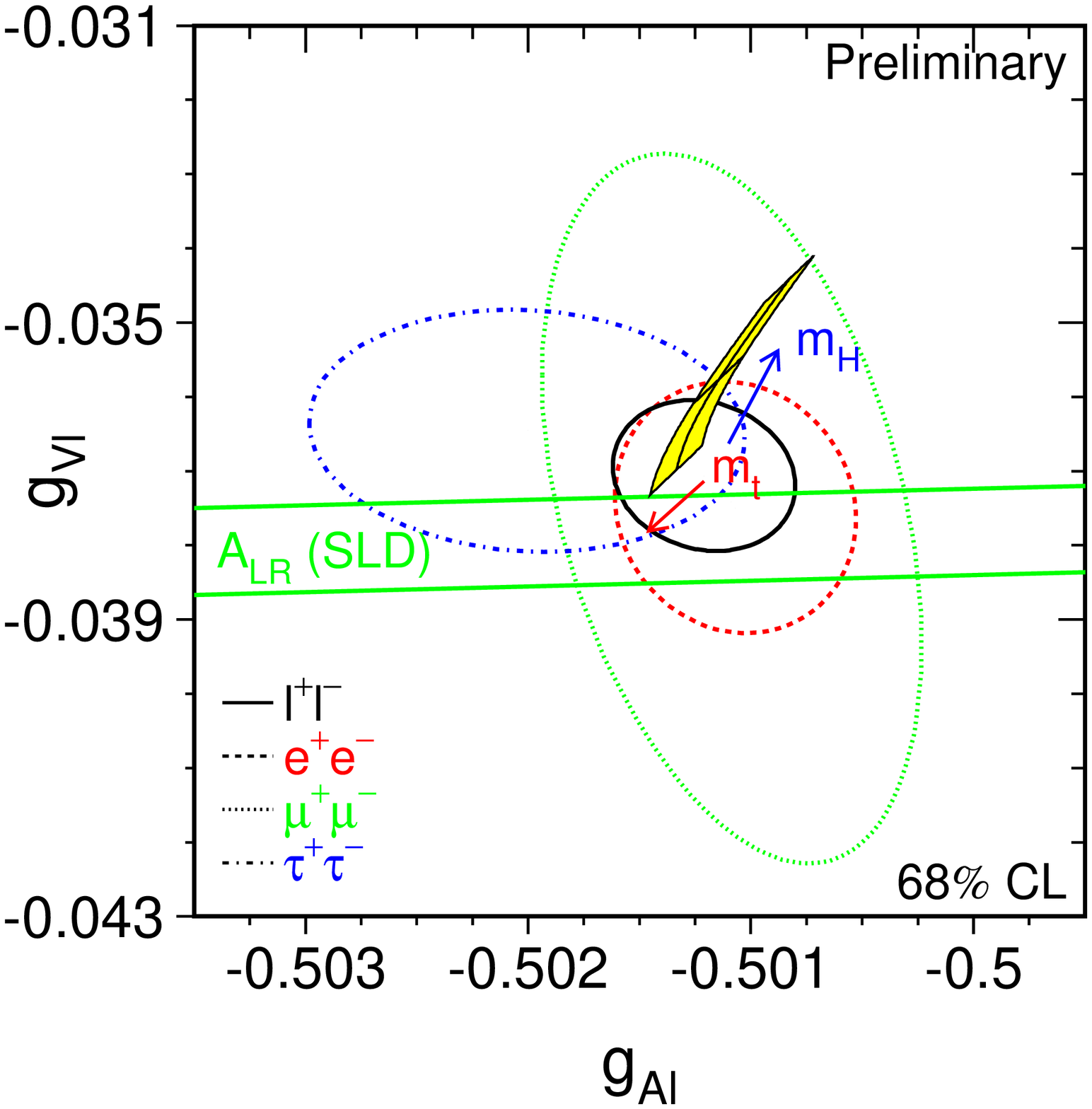}} 
\end{center}
\end{minipage}
\hfill    
\begin{minipage}[c]{.44\linewidth}
\vspace{-\abovedisplayskip}
\caption{68\% probability contours in the $a_l$-$v_l$ plane  from LEP measurements
\protect\cite{LEP:99}. The solid contour assumes lepton universality.   Also shown
is the $1\sigma$ band resulting from the 
$\protect\cA_{\mbox{\protect\rms LR}}^0$ measurement at SLD.   The shaded region
corresponds to the SM prediction  for \protect{$m_t= 174.3\pm 5.1$} GeV and 
\protect{$m_H = 300^{+700}_{-210}$} GeV.  The arrows point in the direction of increasing 
$m_t$ and $m_H$ values.  }
\label{fig:gagv} 
\end{minipage}
}
\end{figure}

Tables~\ref{tab:LEP_asym} and \ref{tab:pol_asym}
show the present experimental results.
The data are in excellent agreement with the SM predictions
and confirm the universality of the leptonic neutral couplings.
\index{lepton universality}
The average of the two $\tau$ polarization measurements, \index{tau polarization}
$\cA_{\mbox{\rms Pol}}^{0,\tau}$ and 
${4\over 3}\cA^{0,\tau}_{\mbox{\rms FB,Pol}}$,  results in 
$\cP_l = -0.1450\pm 0.0033$ which deviates
by $1.5\,\sigma$ from the $\cA^0_{LR}$ measurement.  Assuming lepton
universality,   the combined result from all leptonic asymmetries gives 
\beq
\cP_l = - 0.1497\pm 0.0016  \, .
\eeq{eq:average_P_l}  

Figure~\ref{fig:gagv} shows the  68\% probability contours in the $a_l$--$v_l$ plane, 
obtained from  a combined analysis \cite{LEP:99} of all leptonic observables.  
Lepton universality is now tested to the $0.15\%
$ level for the axial--vector neutral couplings, while only a few per cent 
precision has been achieved for the vector couplings \cite{LEP:99,swartz}:
\beqa {a_\mu / a_e} = 1.0001\pm 0.0014 \, , \qquad &  & 
{v_\mu / v_e} = 0.981\pm 0.082\, ,
\CR
 {a_\tau / a_e} = 1.0019\pm 0.0015  \, , \qquad &  & 
 {v_\tau / v_e} = 0.964\pm 0.032
\, .
\eeqan

Assuming lepton universality, the measured leptonic asymmetries can be 
used to obtain the  effective electroweak mixing angle in the 
charged--lepton sector \cite{LEP:99,swartz}:
\beq 
\sin^2{\theta^{\mbox{\rms lept}}_{\mbox{\rms eff}}} \equiv  {1\over 4}   
\left( 1 - {v_l\over a_l}\right)  \, = \, 0.23119\pm 0.00021 \quad\qquad
(\chi^2/\mbox{d.o.f.} = 3.4/4) .
\eeq{eq:bar_s_W_l} 
%

\section{Lorentz Structure}  
\label{sec:lorentz}

Let us consider the leptonic decay $l^-\to\nu_l l'^-\bar\nu_{l'}$. 
The most general, local, derivative--free, lepton--number conserving, 
four--lepton interaction Hamiltonian, 
consistent with locality and Lorentz invariance
\cite{MI:50,FGJ:86},
\beq
{\cal H} \, =\,  4 \frac{G_{l'l}}{\sqrt{2}}\,
\sum_{n,\epsilon,\omega}  g^n_{\epsilon\omega}\,   
\left[ \overline{l'_\epsilon} 
\Gamma^n {(\nu_{l'})}_\sigma \right]\, 
\left[ \overline{({\nu_l})_\lambda} \Gamma_n 
	l_\omega \right]\ ,
\eeq{eq:hamiltonian}
contains ten complex coupling constants or, since a common phase is
arbitrary, nineteen independent real parameters
which could be different for each leptonic decay.
The subindices
$\epsilon , \omega , \sigma, \lambda$ label the chiralities (left--handed,
right--handed)  of the  corresponding  fermions, and $n$ the
type of interaction: 
scalar ($I$), vector ($\gamma^\mu$), tensor 
($\sigma^{\mu\nu}/\sqrt{2}$).
For given $n, \epsilon ,
\omega $, the neutrino chiralities $\sigma $ and $\lambda$
are uniquely determined.

The total decay width is proportional to the following combination
of couplings, which is usually normalized to one \cite{FGJ:86}:
\beqa\label{eq:normalization}
1 & = &
{1\over 4} \,\left( |g^S_{RR}|^2 + |g^S_{RL}|^2
    + |g^S_{LR}|^2 + |g^S_{LL}|^2 \right)
   +  3 \,\left( |g^T_{RL}|^2 + |g^T_{LR}|^2 \right) 
\\ & &\mbox{}
+ \left(
   |g^V_{RR}|^2 + |g^V_{RL}|^2 + |g^V_{LR}|^2 + |g^V_{LL}|^2 \right)
\CR & \equiv &
Q_{LL} + Q_{LR} + Q_{RL} + Q_{RR} \, .
\eeqan
The universality tests mentioned before refer then to the global
normalization $G_{l'l}$, while the $g^n_{\epsilon \omega}$ couplings
parametrize the relative strength of different types of interaction.
The sums $Q_{\epsilon \omega}$ of all factors with the
same subindices give the probability of having a decay from an
initial charged lepton with chirality $\omega$ to a final one with
chirality $\epsilon$.
In the SM, $g^V_{LL}  = 1$  and all other
$g^n_{\epsilon\omega} = 0 $.

The energy spectrum and angular distribution of the
final charged lepton provides information on
the couplings $g^n_{\epsilon \omega}$.
For unpolarized leptons, the distribution is characterized by
the so-called Michel \cite{MI:50} parameter $\rho$
and the low--energy parameter $\eta$. Two more parameters, $\xi$
and $\delta$, can be determined when the initial lepton polarization is known.
In the SM, $\rho = \delta = 3/4$, $\eta = 0$ and $\xi = 1$.

For $\mu$ decay, where precise measurements of the $\mu$ and $e$
polarizations 
have been performed, there exist \cite{FGJ:86}
upper bounds on $Q_{RR}$, $Q_{LR}$ and $Q_{RL}$, and a lower bound
on $Q_{LL}$. They imply corresponding upper limits on the 8
couplings $|g^n_{RR}|$, $|g^n_{LR}|$ and $|g^n_{RL}|$.
The measurements of the $\mu^-$ and the $e^-$ do not allow to
determine $|g^S_{LL}|$ and $|g^V_{LL}|$ separately;
nevertheless, since the helicity of the $\nu_\mu$ in pion decay is
experimentally known
to be $-1$, a lower limit on $|g^V_{LL}|$ is
obtained from the inverse muon decay
$\nu_\mu e^-\to\mu^-\nu_e$.
These limits show nicely 
that the bulk of the $\mu$--decay transition amplitude is indeed of
the predicted V$-$A type:
$|g^V_{LL}| > 0.960$ \ (90\% CL) \cite{PDG:98}.

\begin{figure}[htb]
\centerline{
\begin{minipage}[c]{.5\linewidth}
\begin{center}
\myfigure{\epsfxsize =7.7cm \epsfbox{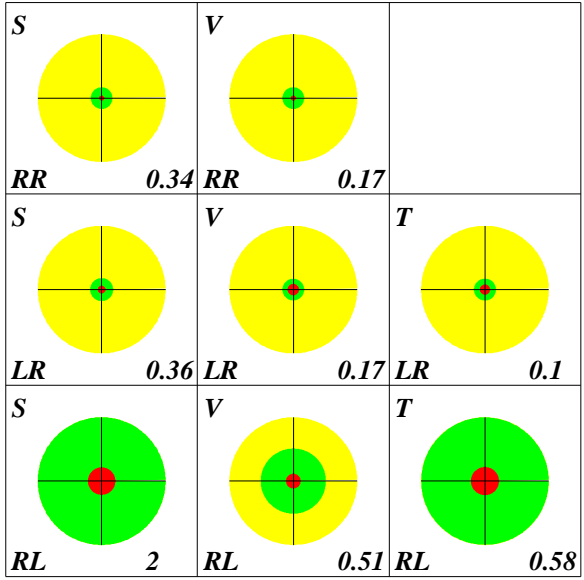}}
\end{center}
\end{minipage}
\hfill
\begin{minipage}[c]{.42\linewidth}
\vspace{-\abovedisplayskip}
\caption{90\% CL experimental limits \protect\cite{Boyko} 
for the normalized $\tau $--decay couplings 
$g'^n_{\epsilon\omega }\equiv g^n_{\epsilon\omega }/ N^n$, where 
$N^n\equiv \mbox{\rm max}(|g^n_{\epsilon\omega }|) = 2, 1, 
1/\protect\sqrt{3}$
for $n=S, V, T$, assuming $e/\mu $ universality.
The circles of unit area indicate the range allowed by the
normalization constraint.
The present experimental bounds are shown as shaded circles.
For comparison, the (stronger) $\mu $--decay limits are also 
shown (darker circles).} 
\label{fig:tau_couplings} 
\end{minipage}
}
\end{figure}

\begin{table}[thb]
\begin{center}
\begin{tabular}{c|c|c|c|c}
& $\mu\to e$ & $\tau\to\mu$ & $\tau\to e$ & $\tau\to l$ 
\\ \hline
$\rho$ & $0.7518\pm 0.0026$ & $0.771\pm 0.025$ & $0.746\pm 0.011$ & 
$0.752\pm 0.009$ 
\\
$\eta$ & $-0.007\pm 0.013\phantom{-}$ & $0.173\pm 0.122$ & --- & 
$0.035\pm 0.031$
\\
$\xi$ & $1.0027\pm 0.0085$ & $1.053\pm 0.053$ & $0.995\pm 0.042$ & 
$0.978\pm 0.031$ 
\\
$\xi\delta$ & $0.7506\pm 0.0074$ & $0.786\pm 0.039$ & $ 0.733\pm 0.029$ & 
$ 0.745\pm 0.021$ 
\\ \hline
\end{tabular}
\caption{World average \protect\cite{Boyko}
Michel parameters. 
The last column ($\tau\to l$) assumes identical couplings
for $l=e,\mu$.}
\label{tab:tau_michel}
\end{center}
\end{table}
%

The experimental analysis of the $\tau$--decay parameters is 
necessarily
different from the one applied to the muon, because of the much
shorter $\tau$ lifetime.
The measurement of the $\tau$ polarization and the parameters
$\xi$ and $\delta$ 
is still possible due to the fact that the spins
of the $\tau^+\tau^-$ pair produced in $e^+e^-$ annihilation 
are strongly correlated \cite{ABGPR:92}.
Another possibility is to use
the beam polarization, as done by SLD.
However,
the polarization of the charged lepton emitted in the $\tau$ decay
has never been measured. 
The measurement of the inverse decay $\nu_\tau l^-\to\tau^-\nu_l$
looks far out of reach.

The four LEP experiments, ARGUS, CLEO and SLD have performed accurate
measurements of the $\tau$--decay Michel parameters. \index{Michel parameters}
The present experimental status \cite{Boyko}
is shown in Table~\ref{tab:tau_michel}.
The determination of the $\tau$ polarization parameters \index{tau polarization}
allows us to bound the total probability for the decay of
a right--handed $\tau$,
\beq\label{eq:Q_R}
Q_{\tau_R} \equiv Q_{RR} + Q_{LR}
= \frac{1}{2}\, \left[ 1 + \frac{\xi}{3} - \frac{16}{9} 
(\xi\delta)\right] \; .
\eeqn
At 90\% CL, one finds (ignoring possible correlations among the measurements):
\beq
Q_{\tau_R}^{\tau\to\mu} < 0.047 \, , \qquad
Q_{\tau_R}^{\tau\to e}  < 0.054 \, , \qquad
Q_{\tau_R}^{\tau\to l}  < 0.032 \, , \qquad
\eeq{eqq:QRlim}
where the last value refers to the $\tau$ decay into either $l=e$ or $\mu$,
assuming identical $e$/$\mu$ couplings.
These probabilities imply corresponding limits on all
$|g^n_{RR}|$ and $|g^n_{LR}|$ couplings. 
Including also the information from $\rho$ and $\eta$,
one gets the  (90\% CL) bounds on the $\tau$--decay couplings shown in
Fig.~\ref{fig:tau_couplings}, where $e$/$\mu$ universality has been assumed.

The probability $Q_{\tau\to l_R}\equiv Q_{RL} + Q_{RR}$
for the $\tau$ decay into a right--handed lepton
could be investigated
through the photon distribution \cite{SV:97}
in the decays $\tau^-\to\nu_\tau l^-\bar\nu_l\gamma$. 
Owing to the large backgrounds, no useful limits can be extracted 
from the recent CLEO measurement of these radiative decays
\cite{CLEOrad}.

\section{Lepton--Number Violation}
\label{sec:LNV}

In the minimal SM with massless neutrinos, there is a
separately conserved additive lepton number for each generation. All
present data are
consistent with this conservation law. However, there are no strong
theoretical      \index{lepton number}
reasons forbidding a mixing among the  different  leptons, in the same
way as happens in the quark sector.
Many models in fact predict lepton--flavour or even
lepton--number violation at some level.
Experimental searches for these processes
can provide information on the scale at which the new physics begins to
play a  significant role.

Table \ref{tab:LVlim} shows the most recent
 limits \cite{CLEOlim} on lepton--flavour and
lepton--number violating decays of the $\tau$.
Although still far
away from the impressive bounds \cite{PDG:98} obtained in $\mu$ decay
[$Br(\mu^-\to e^- \gamma)  < 4.9 \times 10^{-11},
Br(\mu^-\to e^- e^+ e^-) < 1.0 \times 10^{-12},
Br(\mu^-\to e^-\gamma\gamma) <  7.2 \times 10^{-11} \, $ (90\% CL)],
the $\tau$--decay limits start to put interesting constraints
on possible new physics contributions.

\begin{table}[thb]
\begin{center}
\begin{tabular}{c|c||c|c||c|c||c|c}
$X^-$ & U. L. & $X^-$ & U. L. & $X^-$ & U. L. & $X^-$ & U. L.
\\ \hline
$e^-\gamma$ & 2.7 & $\mu^-\gamma$ & 1.1 &
$e^-\pi^+\pi^-$ & 2.2 & $\mu^-\pi^+\pi^-$ & 8.2 \\
$e^-e^+e^-$ & 2.9 & $\mu^-\mu^+\mu^-$ & 1.9 & 
$e^-\pi^+K^-$ & 6.4 & $\mu^-\pi^+K^-$ & 7.5 \\
$e^-e^+\mu^-$ & 1.7 & $\mu^-\mu^+e^-$ & 1.8 &
$e^-K^+\pi^-$ & 3.8 & $\mu^-K^+\pi^-$ & 7.4 \\
$e^-\mu^+e^-$ & 1.5 & $\mu^-e^+\mu^-$ & 1.5 & 
$e^-K^+K^-$ & 6.0 & $\mu^-K^+K^-$ & 15  \\
$e^-\pi^0$ & 3.7 & $\mu^-\pi^0$ & 4.0 &
$e^+\pi^-\pi^-$ & 1.9 & $\mu^+\pi^-\pi^-$ & 3.4 \\
$e^-\eta$ & 8.2 & $\mu^-\eta$ & 9.6 & 
$e^+\pi^-K^-$ & 2.1 & $\mu^+\pi^-K^-$ & 7.0  \\
$e^-\rho^0$ & 2.0 & $\mu^-\rho^0$ & 6.3 &
$e^+K^-K^-$ & 3.8 & $\mu^+K^-K^-$ & 6.0 \\
$e^-K^{*0}$ & 5.1 & $\mu^-K^{*0}$ & 7.5 & 
$e^-\pi^0\pi^0$ & 6.5 & $\mu^-\pi^0\pi^0$ & 14 \\
$e^-\bar K^{*0}$ & 7.4 & $\mu^-\bar K^{*0}$ & 7.5 &
$e^-\pi^0\eta$ & 24  & $\mu^-\pi^0\eta$ & 22 \\
$e^-\phi$ & 6.9 & $\mu^-\phi$ & 7.0 & 
$e^-\eta\eta$ & 35 & $\mu^-\eta\eta$ & 60 
\\ \hline
\end{tabular}
\caption{90\% CL upper limits (in units of $10^{-6}$)
on $B(\tau^-\to X^-)$
\protect\cite{CLEOlim}}
\label{tab:LVlim}
\vspace{-1cm}
\end{center}
\end{table}
%

\section{The Tau Neutrino}

\begin{table}[htb]
\begin{center}
\begin{tabular}{c|c|c|c|c}  
$X$ & ALEPH \cite{ALEPHnumass} & CLEO \cite{CLEOnumass} & 
DELPHI \cite{DELPHInumass} & OPAL \cite{OPALnumass}\\ \hline
$3\pi$ & $25.7$ & --- & $28$ & $35.3$ \\
$3\pi\pi^0$ & --- & $28$ & --- & \\
$5\pi$ & $23.1$ & $30$ & --- & $43.2$ \\ \hline
Combined & $18.2$ & $28$ & $28$ & $27.6$
\\ \hline
\end{tabular}
\caption{95\% CL upper limits on $m_{\nu_\tau}$ (in MeV), from
$\tau^-\to\nu_\tau X^-$ events.}
\label{tab:nulim}
\end{center}
\end{table}

All observed $\tau$ decays are supposed to be accompanied by neutrino
emission, in order to fulfil energy--momentum conservation requirements.
From a two--dimensional likelihood fit of the
visible energy and the invariant--mass distribution of the final hadrons
in $\tau^-\to\nu_\tau X^-$ events, it is possible to set a bound on
the $\nu_\tau$ mass. The best limits, shown in Table~\ref{tab:nulim},
are obtained from modes with a significant probability
of populating the hadronic--mass end--point region. \index{neutrino mass}


The present data are consistent with the $\nu_\tau$
being a conventional sequential neutrino. 
Since taus are not produced
by $\nu_e$ or $\nu_\mu$ beams, we know that $\nu_\tau$
is different from the electronic and  muonic neutrinos.
LEP and SLC have confirmed \cite{LEP:99}
the existence of three (and only
three) different light neutrinos, with standard couplings to the $Z$.
However, no direct observation of $\nu_\tau$, that is, interactions resulting
from $\tau$ neutrinos, has been made so far.
The DONUT experiment at Fermilab is expected to provide soon
the first evidence of $\tau$ neutrinos (produced through
$p+N\to D_s + \cdots $,
followed by the decays $D_s\to\tau^-\bar\nu_\tau$
and $\tau^-\to\nu_\tau + \cdots $), through the detection
of $\nu_\tau + N\to\tau +X$.  
This is an important goal in view of the recent
SuperKamiokande results suggesting
$\nu_\mu\to\nu_\tau$ oscillations with 
$m_{\nu_\tau}^2-m_{\nu_\mu}^2\sim (0.05\;\mbox{\rm eV})^2$.
This hypothesis could be corroborated making
a long--baseline
neutrino experiment with a $\nu_\mu$ beam pointing into a far
($\sim 700$ Km) massive detector, able to detect the appearance of
a $\tau$. The possibility to perform such an experiment
is presently being investigated.

\section{Hadronic Decays}
\label{sec:hadronic}

\begin{figure}[htb]
\centerline{
\begin{minipage}[c]{.5\linewidth}
\begin{center}
\myfigure{\epsfxsize =7.5cm \epsfbox{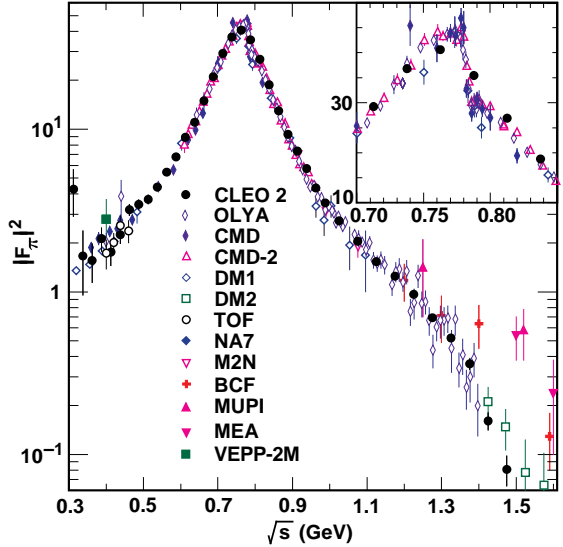}}
\end{center}
\end{minipage}
\hfill
\begin{minipage}[c]{.44\linewidth}
\vspace{-\abovedisplayskip}
\caption{Pion form factor from $\tau^-\to\nu_\tau\pi^-\pi^0$ data
\protect\cite{CLEOpiff} (filled circles), compared with 
$e^+e^-\to\pi^+\pi^-$ measurements.} 
\label{fig:pionff} 
\end{minipage}
}
\end{figure}

The semileptonic decay modes $\tau^-\to\nu_\tau H^-$
probe the  matrix
element of the left--handed charged current between the vacuum and the
final hadronic state $H^-$.
For the two--pion final state, the hadronic matrix element is parametrized
in terms of the so-called pion form factor: \index{pion form factor}
\beq
\langle \pi^-\pi^0| \bar d \gamma^\mu  u | 0 \rangle \equiv
\sqrt{2}\, F_\pi(s)\, \left( p_{\pi^-}- p_{\pi^0}\right)^\mu \, .
\eeq{eq:Had_matrix}
Figure \ref{fig:pionff} shows the recent CLEO measurement \cite{CLEOpiff}
of $|F_\pi(s)|^2$ from $\tau\to\nu_\tau\pi^-\pi^0$ data
(a similar analysis was done previously by ALEPH \cite{ALEPHpiff}).
Also shown is the corresponding determination from
$e^+e^-\to\pi^+\pi^-$ data.
The precision achieved with $\tau$ decays is clearly better.
There is a quite good agreement between both sets of data, although
the $\tau$ points tend to be slightly higher.

The dynamical structure of other hadronic final states has been also 
investigated.
CLEO has measured recently \cite{CLEO3pi}
the four $J^P=1^+$ structure functions
characterizing the decay $\tau^-\to\nu_\tau\pi^- 2\pi^0$,
improving a previous OPAL analysis \cite{OPAL3pi}.
The interference between the two $\pi^-\pi^0$ systems generates
a parity--violating angular asymmetry \cite{KW:84}, which allows
to determine the sign of the $\nu_\tau$ helicity to be $-1$
(the modulus has been precisely measured through the study of
correlated $\tau^+\tau^-$ decays into different final states:
$|h_{\nu_\tau}| =1.0000\pm 0.0057$ \cite{stahl}). \index{neutrino helicity}

\section{QCD Tests}
\label{sec:QCD}


The inclusive character of the total $\tau$ hadronic width
renders possible an accurate calculation of the ratio
\cite{BNP:92,LDP:92a}
\beq
R_\tau \equiv { \Gamma [\tau^- \rightarrow \nu_\tau
  \,\mbox{\rm hadrons}\, (\gamma)] \over
  \Gamma [\tau^- \rightarrow
  \nu_\tau e^- {\bar \nu}_e (\gamma)] } 
  \, = \, R_{\tau,V} + R_{\tau,A} + R_{\tau,S}\, ,
\eeq{eq:r_tau_def}
using analyticity constraints and the operator product expansion.
One can separately compute the contributions associated with
specific quark currents.
Non-strange hadronic decays of the $\tau$ are resolved experimentally
into vector ($R_{\tau,V}$) and axial-vector ($R_{\tau,A}$)
contributions according to whether the
hadronic final state includes an even or odd number of pions.
Strange decays ($R_{\tau,S}$) are of course identified by the
presence of an odd number of kaons in the final state.

The theoretical prediction for $R_{\tau,V+A}$ can be expressed as 
\beq
R_{\tau,V+A} = N_C\, |V_{ud}|^2\,
S_{EW} \left( 1 + \delta_{EW}'   + \delta_P   + \delta_{NP} 
\right) ,
\eeq{eq:r_total}
with $N_C=3$ the number of quark colours.
The factors $S_{EW}=1.0194$ and $\delta_{EW}'=0.0010$ 
contain the known electroweak corrections at the leading
\cite{MS:88} and next-to-leading \cite{BL:90} logarithm
approximation.
The dominant correction ($\sim 20\% $)
is the purely perturbative QCD contribution \cite{BNP:92,LDP:92a}
\beq\label{eq:delta0}
\delta_P  =
a_\tau + 5.2023 \, a_\tau^2 + 26.366 \, a_\tau^3
       + \, \O(\alpha_s^4)  \, .
\eeqn
This expansion in powers of $a_\tau\equiv\alpha_s(m_\tau^2)/\pi$ 
has rather large coefficients, which originate in the long running
of the strong coupling along a contour integration in the complex plane; 
this running effect can be properly resummed to all orders
in $\alpha_s$ by fully keeping \cite{LDP:92a}
the known four--loop--level calculation of the contour integral.

The non-perturbative contributions can be shown to be
suppressed by six powers of the $\tau$ mass \cite{BNP:92},
and are therefore very small. Their actual numerical size has been
determined from the invariant--mass distribution of the final hadrons 
in $\tau$ decay, through the study of weighted integrals \cite{LDP:92b},
\beq
R_{\tau,V+A}^{kl} \equiv \int_0^{m_\tau^2} ds\, 
\left(1 - {s\over m_\tau^2}\right)^k\, \left({s\over m_\tau^2}\right)^l\,
{d R_{\tau,V+A}\over ds} \, ,
\eeq{eq:moments}
which can be calculated theoretically in the same way as $R_{\tau,V+A}$.
The predicted suppression \cite{BNP:92}
of the non-perturbative corrections has been confirmed by
ALEPH \cite{ALEPH:98}, CLEO \cite{CLEO:95} and OPAL \cite{OPAL:98}.
The most recent analyses \cite{ALEPH:98,OPAL:98} give
\beq
\delta_{\mbox{\rms NP}}  =
-0.003\pm 0.003 \, .
\eeq{eq:del_np}
%

\begin{figure}[htb]
\centerline{
\begin{minipage}[c]{.5\linewidth}
\begin{center}
\myfigure{\epsfxsize =7.5cm\epsfbox{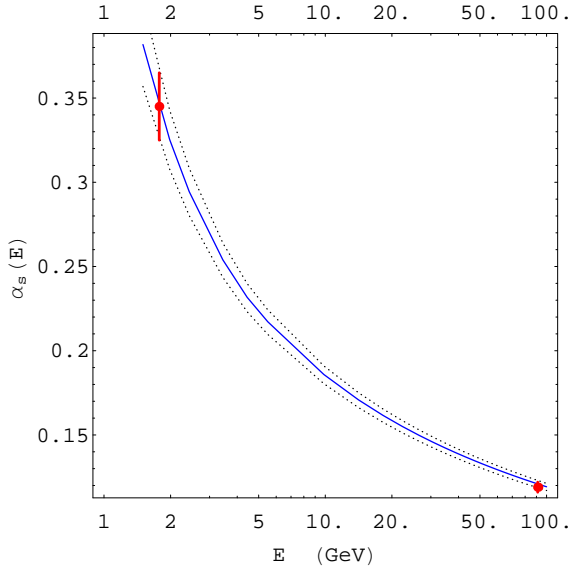}}
\end{center}
\end{minipage}
\hfill
\begin{minipage}[c]{.44\linewidth}
\vspace{-\abovedisplayskip}
\caption{Measured values of $\alpha_s$ in $\tau$ and $Z$ decays.
The curves show the energy dependence predicted by QCD, using
$\alpha_s(m_\tau)$ as input.}
\label{fig:alpha_s}
\end{minipage}
}
\end{figure}

The QCD prediction
for $R_{\tau,V+A}$ is then completely dominated by the
perturbative contribution $\delta_P$; non-perturbative effects being smaller
than the perturbative uncertainties from uncalculated higher--order
corrections. The result turns out to be
very sensitive to the value of $\alpha_s(m_\tau^2)$, allowing for an accurate
determination of the fundamental QCD coupling.
The experimental measurement \cite{ALEPH:98,OPAL:98} 
$R_{\tau,V+A}= 3.484\pm0.024$ implies $\delta_P = 0.200\pm 0.013$,
which corresponds (in the $\overline{\rm MS}$ scheme) to 
\beq\label{eq:alpha}
\alpha_s(m_\tau^2)  =  0.345\pm0.020 \, . 
\eeqn

The strong coupling measured at the $\tau$ mass
scale is significatively different from the values obtained at higher energies.
From the hadronic decays of the $Z$ boson, one gets
$\alpha_s(M_Z) = 0.119\pm 0.003$,  which differs from the $\tau$ decay
measurement by eleven standard deviations!
After evolution up to the scale $M_Z$ \cite{Rodrigo:1998zd}, 
the strong coupling constant 
\index{strong coupling constant} 
in \leqn{eq:alpha} decreases to 
\beq\label{eq:alpha_z}
\alpha_s(M_Z^2)  =  0.1208\pm 0.0025 \, ,
\eeqn
in excellent agreement with the direct measurements at the $Z$ peak
and with a similar accuracy.
The comparison of these two determinations of $\alpha_s$ in two extreme
energy regimes, $m_\tau$ and $M_Z$, provides a beautiful test of the
predicted running of the QCD coupling;
i.e. a very significant experimental verification of {\it asymptotic freedom}.
\index{asymptotic freedom}

\begin{figure}[bht]
\centerline{
\begin{minipage}[b]{.45\linewidth}
\begin{center}
\myfigure{\epsfysize =7cm \epsfbox{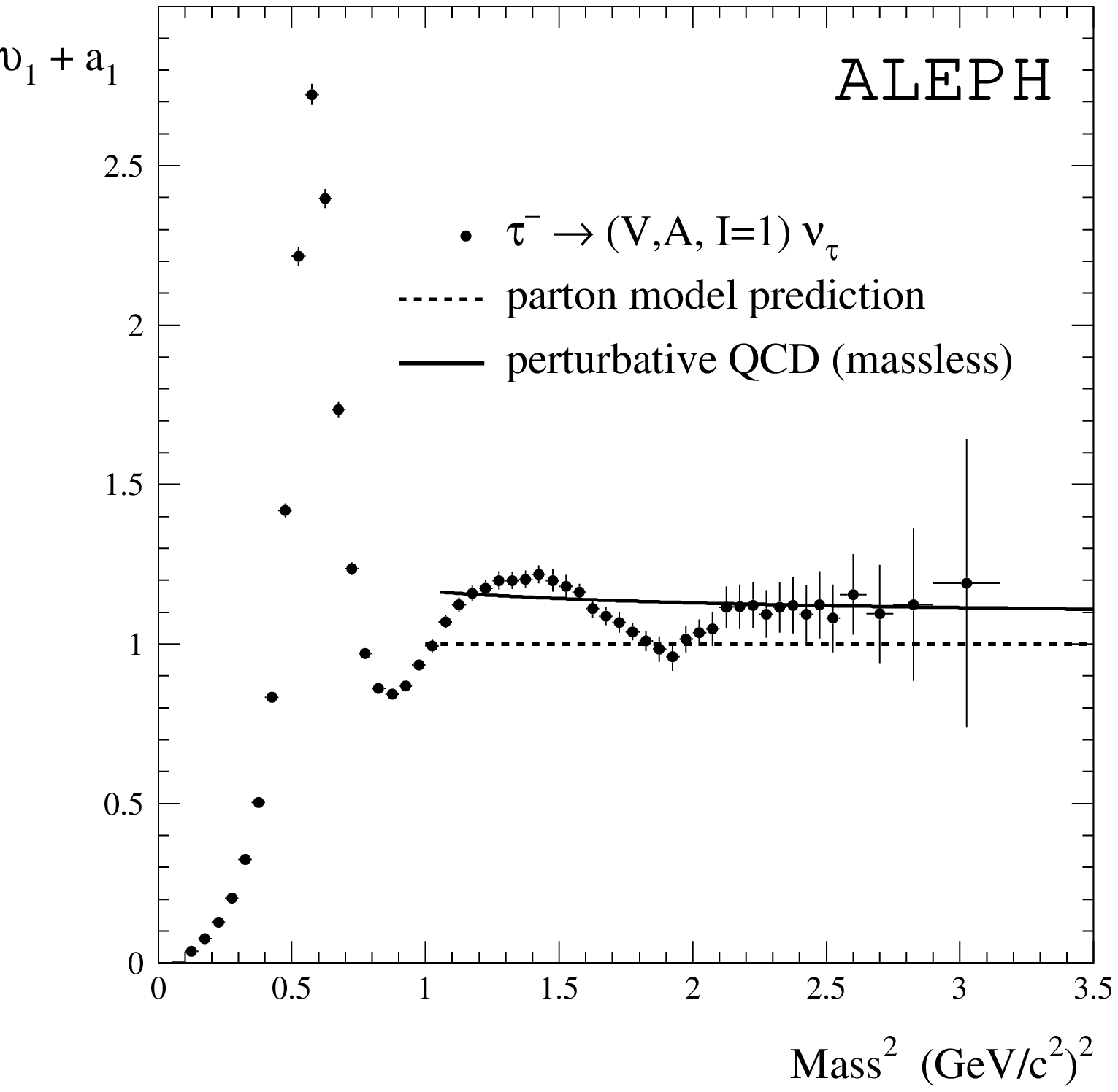}} 
\caption{$V+A$ spectral function \protect\cite{ALEPH:98}.}
\label{fig:V+Asf}
\end{center}
\end{minipage}
\hfill   
\begin{minipage}[b]{.46\linewidth}
\begin{center}
\myfigure{\epsfysize =7cm \epsfbox{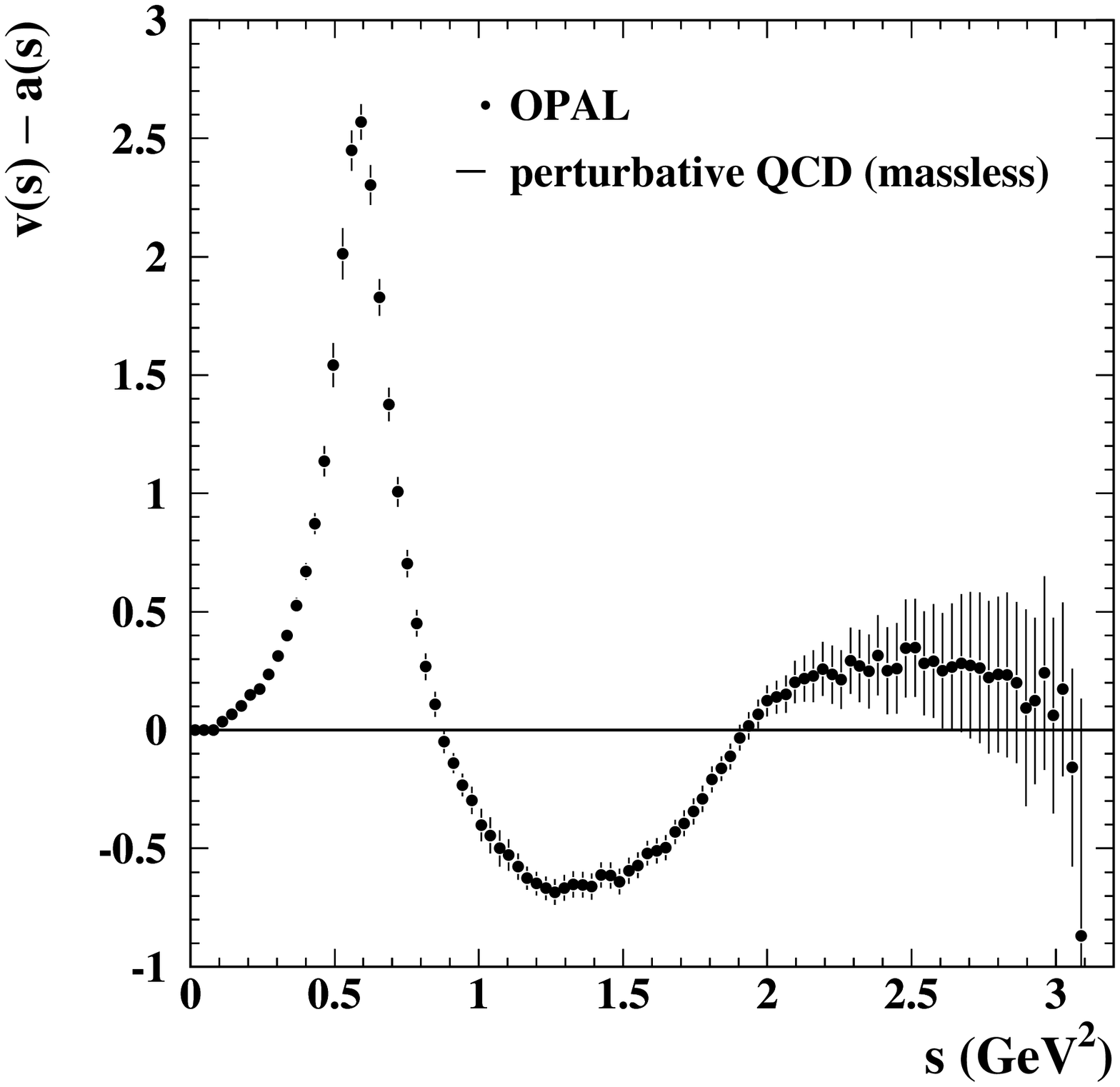}} 
\caption{$V-A$ spectral function \protect\cite{OPAL:98}. }
\label{fig:V-Asf} 
\end{center}
\end{minipage}
}
\end{figure}

From a careful analysis of the hadronic invariant--mass distribution,
ALEPH \cite{ALEPHpiff,ALEPH:98} and OPAL \cite{OPAL:98} have measured
the spectral functions associated with the vector and axial--vector
quark currents. Their difference is a pure non-perturbative quantity,
which carries important information on the QCD dynamics;
it allows to determine low--energy parameters, such
as the pion decay constant, the electromagnetic pion mass difference
$m_{\pi^\pm}-m_{\pi^0}$, or the axial pion form factor,  
in good agreement with their direct measurements.

The vector spectral function has been also used to measure the hadronic
vacuum polarization \index{vacuum polarization}
effects associated with the photon and, therefore, estimate
how the electromagnetic fine structure constant
gets modified at LEP energies.
The uncertainty of this parameter is one
of the main limitations in the extraction of the Higgs mass from global
electroweak fits to the LEP/SLD data.
From the ALEPH $\tau$ data \cite{ALEPHpiff}, 
the Orsay group obtains \cite{orsay}
$\alpha^{-1}(M_Z) = 128.933 \pm 0.021$, which reduces the error
of the fitted $\log{(M_H)}$ value by 30\%.
The same $\tau$ data allows to pin down the hadronic contribution to
the anomalous magnetic moment of the muon \index{anomalous magnetic moment}
$a^\gamma_\mu$. The recent       
analyses \cite{CLEOpiff,orsay} have improved the theoretical
prediction of $a^\gamma_\mu$,
setting a reference value  to be compared with the
forthcoming measurement of the
BNL-E821 experiment, presently running at Brookhaven.

\section{The Strange Quark Mass}

\begin{figure}[htb]
\centerline{
\begin{minipage}[c]{.48\linewidth}
\begin{center}
\myfigure{\epsfxsize =7.5cm \epsfbox{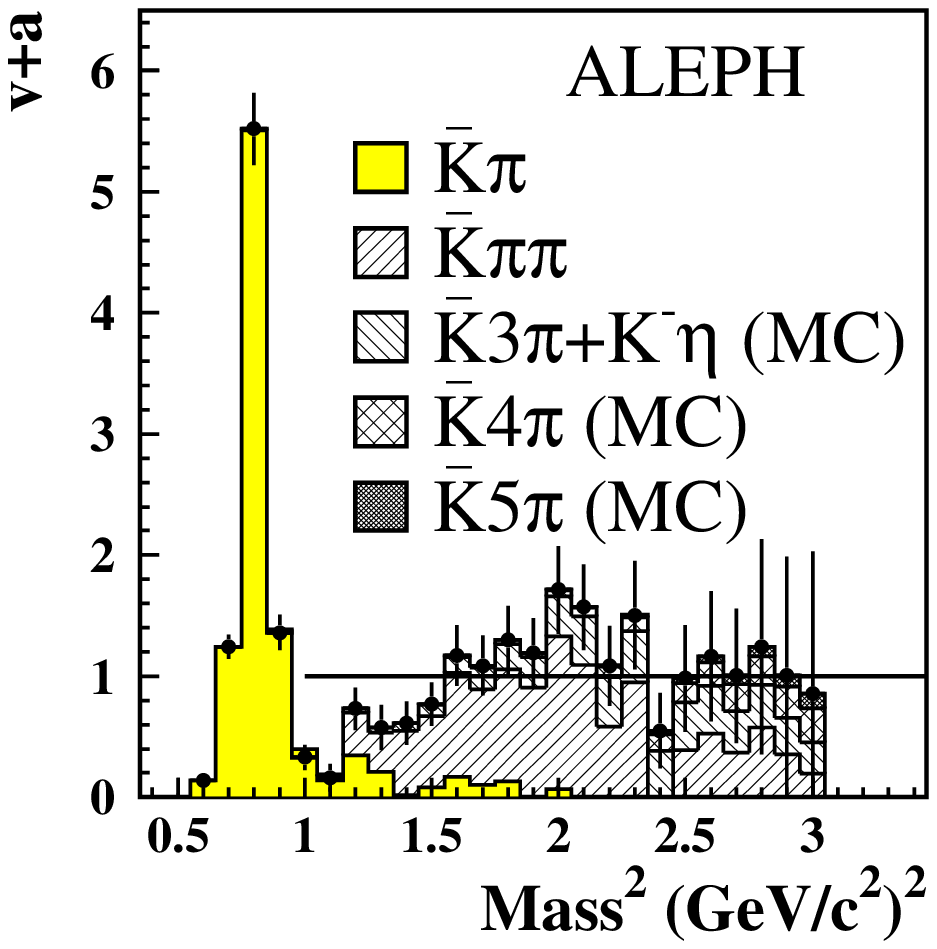}} 
\caption{$|\Delta S|=1$ spectral function \protect\cite{ALEPHms}.}
\label{fig:S=1sf}
\end{center}
\end{minipage}
\hfill
\begin{minipage}[c]{.5\linewidth}
\begin{center}
\begin{tabular}{c|c|c}
$(k,l)$ & $\delta R_\tau^{kl}$ & $m_s(m_\tau)$ \ (MeV)
\\ \hline
$(0,0)$ & $0.394\pm 0.137$ & $143\pm 31_{\rms exp}\pm 18_{\rms th}$
\\
$(1,0)$ & $0.383\pm 0.078$ & $121\pm 17_{\rms exp}\pm 18_{\rms th}$
\\
$(2,0)$ & $0.373\pm 0.054$ & $106\pm 12_{\rms exp}\pm 21_{\rms th}$
\end{tabular}
\vskip 10pt 
Table 8: Measured moments $\delta R_\tau^{kl}$
\protect\cite{ALEPHms}  and corresponding $m_s(m_\tau)$ values
\protect\cite{PP:99}. \hfill\hfil\mbox{}
\end{center}
\end{minipage}
}
\end{figure}

The LEP experiments and CLEO have performed an extensive investigation
of kaon production in $\tau$ decays.
ALEPH has determined the inclusive invariant mass
distribution of the final hadrons in the Cabibbo--suppressed decays 
\cite{ALEPHms}.
The separate measurement of the $|\Delta S|=0$ and $|\Delta S|=1$ 
decay widths allows us to pin down the SU(3) breaking effect induced 
by the strange quark mass, \index{strange quark mass}
through the differences
\beq
\delta R_\tau^{kl}  \equiv
  {R_{\tau,V+A}^{kl}\over |V_{ud}|^2} - {R_{\tau,S}^{kl}\over |V_{us}|^2}
 \,\approx\, 24\, {m_s^2(m_\tau)\over m_\tau^2} \, \Delta_{kl}(\alpha_s)
    - 48\pi^2\, {\delta O_4\over m_\tau^4} \, Q_{kl}(\alpha_s)
\, ,
\eeqn
where $\Delta_{kl}(\alpha_s)$ and $Q_{kl}(\alpha_s)$ are perturbative QCD
corrections, which are known to $O(\alpha_s^3)$ and $O(\alpha_s^2)$,
respectively \cite{PP:99}. 
The small non-perturbative contribution,
$\delta O_4 \equiv\langle 0| m_s \bar s s - m_d \bar d d |0\rangle 
 = -(1.5\pm 0.4)\times 10^{-3}\;\mbox{\rm GeV}^4$,
has been estimated with Chiral Perturbation Theory techniques \cite{PP:99}.
Table~8  
shows the measured \cite{ALEPHms} differences 
$\delta R_\tau^{kl}$ and the corresponding ($\overline{\rm MS}$)
values \cite{PP:99} of $m_s(m_\tau)$. 
The theoretical errors are dominated by the
large perturbative uncertainties of $\Delta_{kl}(\alpha_s)$.
Taking into account the information from the three moments, one finally
gets \cite{PP:99}
\beq
m_s(m_\tau) = \left(119 \pm 12_{\rms exp}\pm 18_{\rms th} \pm 10_{V_{us}}
\right)\;\mbox{\rm MeV} \, ,
\eeqn
where the additional error reflects the present uncertainty from $|V_{us}|$.
This corresponds to
$m_s(1\:\mbox{\rm GeV}^2) = (164\pm 33)$ MeV.

\section{Summary}
\label{sec:summary}

The flavour structure of the SM is one of the main pending questions
in our understanding of weak interactions. Although we do not know the
reason of the observed family replication, we have learned experimentally
that the number of SM fermion generations is just three (and no more).
Therefore, we must study as precisely as possible the few existing flavours
to get some hints on the dynamics responsible for their observed structure.

The $\tau$ turns out to be an ideal laboratory to test the SM. 
It is a lepton, which means clean physics, and moreover it is
heavy enough to produce a large variety of decay modes.
Na\"{\i}vely, one would expect the $\tau$ to be much more sensitive
than the $e$ or the $\mu$ to new physics related to the flavour and
mass--generation problems.

QCD studies can also benefit a lot from the existence of this heavy lepton,
able to decay into hadrons. Owing to their semileptonic character, the
hadronic $\tau$ decays provide a powerful tool to investigate the low--energy
effects of the strong interactions in rather simple conditions.

Our knowledge of the $\tau$ properties has been considerably
improved during the last few years. 
Lepton universality has been tested to rather good accuracy,
both in the charged and neutral current sectors. The
Lorentz structure of the leptonic $\tau$ decays is certainly not determined,
but begins to be experimentally explored.
An upper limit of 3.2\% (90\% CL) has been already set on the probability
of having a (wrong) decay from a right--handed $\tau$.
The quality of the hadronic data
has made possible to perform quantitative QCD tests
and determine the strong coupling constant very accurately.
Moreover, the Cabibbo--suppressed decay width
of the $\tau$ provides a rather good measurement of the strange quark mass.
Searches for non-standard phenomena have been pushed to the limits 
that the existing data samples allow to investigate.

At present, all experimental results on the $\tau$ lepton are consistent with
the SM. There is, however, large room for improvements. Future $\tau$ experiments
will probe the SM to a much deeper level of sensitivity and will explore the
frontier of its possible extensions.

\bigskip
This work has been supported in part by the ECC, TMR Network EURODAPHNE
(ERBFMX-CT98-0169), and by DGESIC (Spain) under grant No. PB97-1261.



\def\Discussion{
\setlength{\parskip}{0.3cm}\setlength{\parindent}{0.0cm}
  \bigskip\bigskip    {\Large {\bf Discussion}} \bigskip }
\def\speaker#1{{\bf #1:}\ }

\Discussion

\speaker{Michel Davier (LAL, Orsay)}
The value you quoted for $m_s(m_\tau)$ from the ALEPH analysis is not the
ALEPH result. Our analysis is more conservative and quotes considerably
larger errors.

\speaker{Pich}
ALEPH \cite{ALEPHms} quotes two different results:
$ (149 {}^{+24_{\rm exp}}_{-30_{\rm exp}} 
 {}^{+21_{\rm th}}_{-25_{\rm th}} \pm 6_{\rm fit})$ MeV
\ and \ 
$(176 {}^{+37_{\rm exp}}_{-48_{\rm exp}} 
 {}^{+24_{\rm th}}_{-28_{\rm th}} \pm 8_{\rm fit} 
\pm 11_{J=0})$ MeV.
The first number is obtained truncating the longitudinal perturbative series
$\Delta_{kl}^L(\alpha_s)$ at $O(\alpha_s)$. The known 
$O(\alpha_s^2)$ and $O(\alpha_s^3)$ contributions are positive; thus, 
you took a 
smaller value of $\Delta_{kl}(\alpha_s)$ and, therefore, got 
a larger result for $m_s$. Your error is larger to account
for the missing  $O(\alpha_s^2)$ and  $O(\alpha_s^3)$ corrections.

The second number has been obtained subtracting the $J=0$ contribution;
unfortunately, only the pion and kaon contributions are known.
Since the longitudinal spectral functions are positive definite,
this procedure gives an upper bound on $m_s$ \cite{PP:99}.
ALEPH 
makes a tiny ad-hoc correction to account for the remaining 
unknown $J=L$ contribution, and quotes the resulting number 
as a $m_s(m_\tau)$ determination. 
Since you added a generous uncertainty, your number does not 
disagree with ours. However, it is actually an 
upper bound on $m_s(m_\tau)$ and not a determination of this 
parameter.

\end{document}